\magnification=1100
\parskip=10pt plus 5pt
\parindent=15pt
\baselineskip=13pt
\input amssym.def
\input amssym
\pageno=0
\footline={\ifnum \pageno < 1 \else \hss \folio \hss \fi}
\topglue 1in
\centerline{\bf The quantum theory of a quadratic gravity action}
\centerline{\bf for heterotic strings}
\vskip .5in
\centerline{\bf Simon Davis \footnote{*}{Present address: 
Institut f{\"u}r Mathematik, Universit{\"a}t Potsdam, 
\hfil\break
\phantom{....} D-14415 Potsdam, Germany}
and Hugh Luckock}
\vskip .5in
\centerline{School of Mathematics and Statistics}
\centerline{University of Sydney}
\centerline{NSW 2006, Australia}
\vskip .5in
\noindent
{\bf Abstract}.  The wave function for the quadratic gravity theory derived
from the heterotic string effective action is deduced to first order in 
${{e^{-\Phi}}\over {g_4^2}}$ by solving a perturbed second-order 
Wheeler-DeWitt equation, assuming that the potential is slowly varying 
with respect to $\Phi$.  Predictions for inflation based on the solution
to the second-order Wheeler-DeWitt equation continue to hold for this
higher-order theory.  It is shown how formal expressions for the 
average paths in minisuperspace 
$\{\langle a(t)\rangle,\langle \Phi(t)\rangle\}$ for this theory can be used 
to determine the shifts from the classical solutions $a_{cl}(t)$ and 
$\Phi_{cl}(t)$, which occur only at third order in the expansion of the 
functional integrals representing the expectation values.

\vfill\eject

Given a dimensionless metric and scalar field, the form of a sigma model 
coupled to gravity [1], renormalizable in the generalized sense, is 
$$\eqalign{I~=~\int~d^4x~{\sqrt{-g}}~&\biggl[b_1(\phi)(\square\phi)^2+b_2(\phi)
(\nabla_\mu\phi)(\nabla^\mu\phi)\square\phi+b_3(\phi)[(\nabla_\mu\phi)
(\nabla^\mu\phi)]^2
\cr
&~+~b_4(\phi)(\nabla_\mu\phi)(\nabla^\mu\phi)
~+~b_5(\phi)~+~c_1(\phi)R(\nabla_\nu\phi)(\nabla^\mu\phi)
\cr
&~+~c_2(\phi)R^{\mu\nu}
(\nabla_\mu\phi)(\nabla_\nu\phi)
~+~c_3(\phi)R\square\phi+a_1(\phi)R_{\mu\nu\alpha\beta}R^{\mu\nu\alpha\beta}
\cr
&~+a_2(\phi)R_{\mu\nu}R^{\mu\nu}
~+~a_3(\phi)R^2~+~a_4(\phi)R\biggr]~+~surface~terms
\cr}
\eqno(1)
$$
where $[b_4(\phi)]=2$, $[b_5(\phi)]=4$ and $[a_4(\phi)]=2$, with 
$a_i(\phi)\ne 0$ for at least one $i\in \{1,2,3\}$ and a renormalizable
potential term $b_5(\phi)$.
Setting
$$\eqalign{b_1(\phi)&=b_2(\phi)=b_3(\phi)=0
\cr
b_4(\phi)~&=~{1\over {2\kappa^2}}~~~~~~b_5(\phi)=-{1\over {\kappa^2}}
{\tilde V}(\phi)
\cr
c_1(\phi)~&=~c_2(\phi)~=~c_3(\phi)=0
\cr
a_1(\phi)~&=~{{e^{{-\phi}\over \kappa}}\over {4g_4^2}}~~~
a_2(\phi)~=~-{{e^{{-\phi}\over \kappa}}\over {g_4^2}}~~~
a_3(\phi)~=~{{e^{{-\phi}\over \kappa}}\over {4g_4^2}}~~~
a_4(\phi)~=~{1\over {\kappa^2}}
\cr} 
\eqno(2)
$$
produces an action describing an exponential coupling of a scalar field
to quadratic curvature terms.  Defining the dilaton field to have 
dimension 1,
$\Phi={\phi\over \kappa}$, and $V(\Phi)~=~{1\over {\kappa^2}}{\tilde V}(\phi)$,
the quadratic gravity theory    
$$I=\int~d^4x~{\sqrt {-g}}\left[{1\over {\kappa^2}}R+{1\over 2}(D\Phi)^2+
{{e^{-\Phi}}\over {4g_4^2}}\times (R_{\mu\nu\kappa\lambda}
R^{\mu\nu\kappa\lambda}-4R_{\mu\nu}R^{\mu\nu}+R^2)-V(\Phi)\right]
\eqno(3)
$$
is equivalent to the one-loop heterotic string effective action with the 
coefficient of the $R{\tilde R}$ term set to zero.  A non-minimal coupling 
$\xi R \phi^2$ term, which has been found useful for generating 
open inflationary universe models [2], can be added to this action  
by setting the coefficient $a_4(\phi)$ equal to ${1\over {\kappa^2}}
(1+\xi \phi^2)$. 

A conformal transformation of the $R^2$ term produces a Ricci scalar plus 
an extra scalar field, and any factor multiplying $R$ can be eliminated
through another conformal transformation [3-5].  There is also a Legendre
transformation of the $R^{\mu\nu}R_{\mu\nu}$ term to a Ricci scalar together
with extra tensor modes [6-8].  The scalar and tensor modes contain an 
intricate dependence on the derivatives of the metric and Ricci tensors, 
which complicates the definition of the corresponding momenta, so that there 
is no simplification in the canonical quantization procedure by applying a 
conformal or Legendre transformation to the action.  The 
$R^{\mu\nu\rho\sigma}R_{\mu\nu\rho\sigma}$ can be rewritten as a linear 
combination of $C^{\mu\nu\rho\sigma}C_{\mu\nu\rho\sigma}$ and terms 
containing $R_{\mu\nu}$ and $R$.  The renormalizability of the action with 
$C^{\mu\nu\rho\sigma}C_{\mu\nu\rho\sigma}$ has been established [9].  Its 
effect on unitarity must be counterbalanced by that of the other tensor modes.

For an action with a polynomial coupling between the scalar field and the
quadratic curvature term $a_1(\phi),~a_2(\phi)$ and $a_3(\phi)$ will
be shifted to ${\tilde a}_1(\phi)$, ${\tilde a}_2(\phi)$ and ${\tilde
a}_3(\phi)$.  The coefficients in a truncated Taylor series representation
of ${{e^{-\Phi}}\over {g_4^2}}$, $\sum_{n=0}^{N_0}~{1\over {n!}}
{{(-1)^n}\over {g_4^{2n}}}\Phi^n+{1\over {(N_0+1)!}}(-1)^{N_0+1}
{{e^{-\lambda\Phi}}\over {g_4^2}}$, $\lambda<1$, would be shifted so that
the couplings of the renormalized theory are
$\sum_{n=0}^{N_0^\prime}~{1\over {n!}}c_{i,n}\Phi^n+
{1\over {(N_0!+1)!}}{\tilde a}_i^{(N_0^\prime+1)}(\lambda^\prime \Phi),
~\lambda^\prime <1$.  The remainder term can be made arbitrarily small
and the series converges as $N_0^\prime\to \infty$ unless $c_{i,n}$ has 
a factorial dependence on $n$, so that the new action with coefficients
${\tilde a}_i(\Phi),~i=1,2,3$, will be defined and renormalizable in 
the generalized sense.

After imposing the restriction to the minisuperspace of 
Friedmann-Robertson-Walker metrics, and adding a boundary term to the
action to eliminate terms containing ${\ddot a}$, the one-dimensional action
is

$$\eqalign{I~&=~\int~dt\biggl[6a(-{\dot a}^2+K)~+~{1\over 2}a^3{\dot \Phi}^2
\cr
&~~~~~~~~~~~~~~~~~+~{{e^{-\Phi}}\over {g_4^2}}{\dot \Phi}{\dot a}
({\dot a}^2+3K)~-~a^3V(\Phi)\biggr].
\cr}
\eqno(4)
$$
where $a(t)$ is the scale factor of the Friedmann-Robertson-Walker universe,
which is open, flat or closed if $K=-1,0$ or $1$.

Given the conjugate momenta 
$$\eqalign{P_a~&=~-12a{\dot a}+6{{e^{-\Phi}}\over {g_4^2}}{\dot \Phi}
({\dot a}^2+K)
\cr
P_\Phi~&=~a^3{\dot \Phi}~+~2{{e^{-\Phi}}\over {g_4^2}}{\dot a}({\dot a}^2
+3K)
\cr}
\eqno(5)
$$
the Wheeler-De Witt 
equation to first order in ${{e^{-\Phi}}\over {g_4^2}}$ [10] is
$$\eqalign{H\Psi~&=~(H_0+{{e^{-\Phi}}\over {g_4^2}}H_1)\Psi~=~0
\cr
H_0~&=~{1\over {24}}{\partial\over {\partial a}}{1\over a}
{\partial\over {\partial a}}~-~{1\over {2a^3}}{{\partial^2}\over 
{\partial\Phi^2}}~-~6aK~+~a^3V(\Phi)
\cr
H_1~&=~{1\over {a^4}}\left({K\over 4}-{1\over {576a^4}}\right){\partial\over
{\partial a}}~+~{1\over {576a^7}}{{\partial^2}\over {\partial a^2}}
~-~{1\over {1728a^6}}{{\partial^3}\over {\partial a^3}}
\cr
&~~~~+~{1\over {a^5}}\left({{7K}\over 4}-{{35}\over {576a^4}}\right)
{\partial\over {\partial\Phi}}~+~{1\over {24a^8}}{{\partial^2}\over
{\partial a \partial\Phi}}~-~{1\over {64a^7}}{{\partial^3}\over {\partial a^2
\partial\Phi}}~+~{1\over {864a^6}}{{\partial^4}\over 
{\partial a\partial\Phi}}.
\cr}
\eqno(6)
$$
Given that $\Psi\doteq \Psi_0+{{e^{-\Phi}}\over {g_4^2}}\Psi_1$, when
$\vert V^{-1}V^\prime(\Phi)\vert \ll 1$ and $\vert V(\Phi)\vert \ll 1$
in Planck units, so that the derivatives of the quantum cosmological wave
function $\Psi$ with respect to $\Phi$ are negligible, the equation
becomes $H_0\Psi_1\approx -H_1 \Psi_0$, to first order in ${{e^{-\Phi}}\over
{g_4^2}}$.  It may be noted that $H_0\left({{e^{-\Phi}}\over {g_4^2}}\Psi_1
\right)$ would contain a term of the form ${1\over {2a^3}}{{\partial^2}\over
{\partial \Phi^2}}\left({{e^{-\Phi}}\over {g_4^2}}\Psi_1\right)$ giving
rise to a contribution ${{e^{-\Phi}}\over {g_4^2}}{1\over {2a^3}}\Psi_1
-{1\over {a^3}}{{e^{-\Phi}}\over {g_4^2}}{{\partial \Psi_1}\over 
{\partial \Phi}}+{1\over {2a^3}}{{e^{-\Phi}}\over {g_4^2}}
{{\partial^2\Psi_1}\over {\partial \Phi^2}}$ to the differential equation at
first order in the expansion parameter.  However, if this term is included, 
then the solution to the standard Wheeler-DeWitt equation
with the ${{\partial^2}\over {\partial\Phi^2}}$ operator and each of the
$\Phi$ derivative terms in $H_1$ would have to be used.  When the
$\Phi$ derivative terms are discarded from at the beginning of the
computation, it is sufficient to consider the differential equation for
$\Psi_1$ without the additional term.

The correction to the wave function is then given by ${{e^{-\Phi}}\over 
{g_4^2}}\Psi_1$ where
$$\Psi_1~\approx~C_1\Psi_{01}~+~C_2\Psi_{02}~-~\Psi_{02}\int~
         \Psi_{01}{{H_1\Psi_0}\over W}a~da~+~24\Psi_{01}\int~\Psi_{02}
                                               {{H_1\Psi_0}\over W}a~da
\eqno(7)
$$
with the Wronskian defined to be
$$W~=~\Psi_{01}{d\over {da}}\Psi_{02}~-~\Psi_{02}{d\over {da}}\Psi_{01}.
\eqno(8)
$$

Consistency with symmetries of the theory depends on the choice of  
boundary condition, which also determines the feasibililty of obtaining
an inflationary cosmology.  Upon consideration of the N=1 supergravity 
theory restricted to the minisuperspace of Bianchi IX metrics, for example,
the requirement of homogeneity implies a Lie derivative condition on the 
spinor fields, which defines a no-boundary ground state [11].  While the
no boundary wave function is defined to be regular in the limit $a\to 0$,
this property only holds for the tunneling wave function when the operator
ordering parameter $p$ is less than one.

The no-boundary wave function is
$$\Psi_{0NB}~=~{{Ai\left(K\left({{36}\over V}\right)^{2\over 3}
\left(1-{{a^2V}\over {6K}}\right)\right)}\over
{Ai\left(K\left({{36}\over V}\right)^{2\over 3}\right)}}
\eqno(9)
$$
when $K=-1$ or $1$.  While the no-boundary wave function is defined by 
a path integral over compact four-manifolds, leading to the conventional
choice $K=1$, the other values of $K$ are possible if the range of
coordinates in the flat or hyperbolic sections is finite.   
While the probability amplitude defined by the no-boundary wave function 
with a positive coefficient in the exponential prefactor does not 
directly imply the existence of an inflationary universe 
with the appropriate e-folding factor, it may be noted that a negative 
coefficient can be obtained by the other choice of sign of 
$(-z_0)^{3\over 2}$ in the asymptotic expansion of $Ai(-z_0)$, 
$z_0=z(a=0)=-K\left({{36}\over V}\right)^{2\over 3}$ as $V(\Phi)\to 0$, 
so that as $z_0\to -\infty$, the normalization factor tends to
${1\over {2{\sqrt \pi}}}(-z_0)^{-{1\over 4}}
e^{\mp {2\over 3}(-z_0)^{3\over 2}}$ if $K=1$.  
Even if the coefficient is positive initially, the value of $V(\Phi)$ would 
be driven to zero, so that the change in sign can be obtained by analytic 
continuation in the variable $V$.   The change in sign of the exponential 
prefactor does not affect the regularity of the wave function in the 
$a(t)\to 0$ limit. When $K=-1$,
$Ai(-z_0)\to {1\over {\sqrt \pi}}(z_0)^{-{1\over 4}}
sin~\left[{2\over 3}z_0^{3\over 2}+{\pi\over 4}\right]$, so that the
wave function diverges ${{24}\over V}+{\pi\over 4}$ tends to $n\pi$, 
$n$ integer.

When $K=0$, it is not necessary to rescale $a(t)$ and $V(\Phi)$ to obtain the 
standard form for the second-order Wheeler-DeWitt equation.  Without a term 
proportional to $K$, one definition of $z$ could be
$(2V)^{-{2\over 3}}(a^2V)=2^{-{2\over 3}}a^2V^{1\over 3}$, 
but then $z(a=0)$ would vanish.  If the normalization factor is chosen to be
$Ai(-z_c)$, where $z_c=z(a_c)$, it can be shown that the probability
distribution is peaked at $V=0$ if $a_c\ll 1$.  The change in sign of $z$
across the $V(\Phi)=0$ boundary leads to different asymptotics for the
wave function.  If the positive sign is chosen for the exponent 
in the prefactor $e^{\mp {2\over 3}(-z_c)^{3\over 2}}=
e^{\mp {1\over 3}a_c^3\vert V\vert^{1\over 2}}$, inflation again can be 
obtained.  
     
Given the two independent solutions of the homogeneous second-order 
differential equation $H_0\Psi_0=0$
$$\eqalign{\Psi_{01}~&=~Ai(-z)~~~~~~~~\Psi_{02}~=~Bi(-z)
\cr
z~&=~-K\left({{36}\over V}\right)^{2\over 3}\left(1-{{a^2V}\over {6K}}\right)
\cr}
\eqno(10)
$$
the Wronskian [14] is

$$\eqalign{W~&=~Ai(-z){d\over {da}}Bi(-z)-Bi(-z){d\over {da}}Ai(-z)
~=~{{dz}\over {da}}\left[Ai(-z){d\over {dz}}Bi(-z)-Bi(-z)){d\over {dz}}Ai(-z)
\right]
\cr
&~=~-{{dz}\over {da}}\pi^{-1}
=~{{aV^{1\over 3}}\over {3\pi}}(36)^{2\over 3}
\cr}
\eqno(11)
$$
and
$$\eqalign{\Psi_1~&=~C_1Ai(-z)~+~C_2Bi(-z)
\cr
&~~~-~{{72\pi}\over {Ai\left(K\left({{36}\over V}\right)^{2\over 3}\right)}}
Bi(-z)\cdot\int~{{da}\over {a^3}}\left[KAi^\prime(-z)
~-~{1\over {36}}\left({V\over {36}}\right)^{2\over 3}
Ai^{\prime\prime\prime}(-z)\right] Ai(-z)
\cr
&~~~+~{{72\pi}\over {Ai\left(K\left({{36}\over V}\right)^{2\over 3}\right)}}
Ai(-z)\cdot \int {{da}\over {a^3}}~\left[KAi^\prime(-z)~+~
{1\over {36}}\left({V\over {36}}\right)^{2\over 3}Ai^{\prime\prime\prime}(-z)
\right] Bi(-z).
\cr}
\eqno(12)
$$

From Airy's differential equation, it follows that 
$A^{\prime\prime\prime}(-z)$ can be replaced by $zAi^\prime(-z)+Ai(-z)$
in the integral, giving

$$\eqalign{\Psi_1~&=~C_1Ai(-z)~+~C_2Bi(-z)
\cr
&-{{72\pi Bi(-z)}\over 
{Ai\left(K\left({{36}\over V}\right)^{2\over 3}\right)}}
\cdot \int {{da}\over {a^3}}\bigg\{\left[\left(-K+{1\over {36}}\left(
{V\over {36}}\right)^{2\over 3}z\right)Ai^\prime(-z)+{1\over {36}}
\left({V\over {36}}\right)^{2\over 3}Ai(-z)\right] Ai(-z)\bigg\}
\cr
&+{{72\pi Ai(-z)}
\over {Ai\left(K\left({{36}\over V}\right)^{2\over 3}\right)}}
\cdot \int {{da}\over {a^3}}\bigg\{\left[\left(-K+{1\over {36}}\left(
{V\over {36}}\right)^{2\over 3}z\right)Ai^\prime(-z)+{1\over {36}}
\left({V\over {36}}\right)^{2\over 3}Ai(-z)\right] Bi(-z)\bigg\}.
\cr}
\eqno(13)
$$

Imposing the Hartle-Hawking boundary condition on the corrected
wave function implies that it must have the same form as the
standard wave function, so that the coefficient of $Bi(-z)$ should
vanish.

After changing the integration variable,
${{da}\over {a^3}}={{3V^{5\over 3}}\over {(36)^{2\over 3}K^2}}
\left(1+\left({V\over {36}}\right)^{2\over 3}{z\over K}\right)^{-2}dz$,
the wave function $\Psi_1$ can be obtained by evaluating two integrals of
the form $\int~{{dz}\over {(1+kz)^2}} Ai^\prime(-z)Ai(-z)(1+k^\prime z)$ and
$\int~{{dz}\over {(1+kz)^2}} [Ai(-z)]^2$.

The integral $\int~dz~\sigma(z)~S_\mu(\phi(z))\cdot S_\nu(\psi(z))$ has
the form $[A(z)S_\mu(\phi(z))+BS_{\mu+1}(\phi(z))]$
$S_\nu(\psi(z))$
\hfil\break
$+[C(z)S_\mu(\phi(z))+D(z)S_{\mu+1}(\phi(z))]
S_{\nu+1}(\psi(z))$ when $S_\mu(z)$ is a cylinder function [15] if

$$\eqalign{\sigma(z)~&=~A^\prime(z)~+~\left(\mu{{\phi^\prime(z)}\over
{\phi(z)}}+\nu{{\psi^\prime(z)}\over {\psi(z)}}\right)A(z)+B\phi^\prime(z)
+C\psi^\prime(z) 
\cr
0~&=~B^\prime(z)~+~\left[\nu{{\psi^\prime(z)}\over {\psi(z)}}-
(\mu+1){{\phi^\prime(z)}\over {\phi(z)}}\right]B(z)~+~D\psi^\prime(z)~-~
A\phi^\prime(z)
\cr
0~&=~C^\prime(z)~+~\left[\mu{{\phi^\prime(z)}\over {\phi(z)}}
-(\nu+1){{\psi^\prime(z)}\over {\psi(z)}}\right]C(z)~+~D\phi^\prime(z)
~-~A\psi^\prime(z)
\cr
0~&=~D^\prime(z)~-~\left[(\mu+1){{\phi^\prime(z)}\over {\phi(z)}}
+(\nu+1){{\psi^\prime(z)}\over {\psi(z)}}\right]D(z)~-~B\psi^\prime(z)
~-~C\phi^\prime(z).
\cr}
\eqno(14)
$$

This set of coupled differential equations can be reduced to the $3\times 3$
system
$${d\over {dz}}\left(\matrix{A(z)&
                               \cr
                              B(z)&
                               \cr
                              C(z)&
                               \cr}\right)
~+~M\left(\matrix{A(z)&
                \cr
                B(z)&
                \cr
                C(z)&
                \cr}\right)~=~\left(\matrix{a(z)&
                                             \cr
                                            b(z)&
                                             \cr
                                            c(z)
                                             \cr}
                                              \right)
\eqno(15)
$$
where
$$\eqalign{M~&=~\left(\matrix{(\mu+\nu){{\phi^\prime(z)}\over {\phi(z)}}&
\phi^\prime(z)& \phi^\prime(z)
\cr
-\phi^\prime(z) & -(1+2\mu){{\phi^\prime(z)}\over {\phi(z)}}+{1\over {\mu-\nu}}
\phi(z)\phi^\prime(z) & (\mu+\nu){{\phi^\prime(z)}\over {\phi(z)}}
                            -{1\over {\mu-\nu}}\phi(z)\phi^\prime(z)
\cr
-\phi^\prime(z) & {1\over {\mu-\nu}}\phi(z)\phi^\prime(z)
                                    &
                 (\mu-\nu-1){{\phi^\prime(z)}\over {\phi(z)}}
                     -{1\over {\mu-\nu}}\phi(z)\phi^\prime(z)\cr}\right)
\cr
&~\left(\matrix{a(z)&
              \cr
              b(z)&
               \cr
              c(z)&
               \cr}\right)
~=~
\left(\matrix{\sigma(z)&
               \cr
               (\mu+\nu)\phi^\prime(z)-{1\over {\mu-\nu}}\phi(z)^2\cdot
                                                              \phi^\prime(z)
               \cr
                -{1\over {\mu-\nu}}\phi(z)^2\cdot \phi^\prime(z)
               \cr}\right).
\cr}
\eqno(16)
$$
The solution to this system of differential equations is
$$\eqalign{\left(\matrix{A(z)&
                  \cr
                B(z)&
                  \cr
                 C(z)&
                  \cr}\right)~&=~exp\left(-\int^z~M(z^\prime)dz^\prime\right)
                                       \cdot
                                      \left(\matrix
                                               {\int^z~a(z^\prime)dz^\prime&
                                                    \cr
                                                \int^z b(z^\prime)dz^\prime&
                                                     \cr
                                                \int^z~c(z^\prime)dz^\prime &
                                                     \cr}\right)
\cr
~&=~T^{-1}Texp\left(-\int^z~M(z^\prime)dz^\prime\right)
                                       \cdot
                                      T^{-1}T\left(\matrix
                                               {\int^z~a(z^\prime)dz^\prime&
                                                    \cr
                                                \int^z~b(z^\prime)dz^\prime&
                                                     \cr
                                                \int^z~c(z^\prime)dz^\prime &
                                                     \cr}\right).
\cr}
\eqno(17)
$$
If T is the matrix which diagonalizes $\int^z~M(z^\prime)dz^\prime$, then
$$T e^{\left(-\int^z~M(z^\prime)dz^\prime\right)} T^{-1}
~=~exp\left(-T\int^z~M(z^\prime dz^\prime)T^{-1}\right)
~=~\left(\matrix{e^{-\lambda_1}&0&0
                        \cr
                    0& e^{-\lambda_2}& 0
                         \cr
                    0& 0& e^{-\lambda_3}
                         \cr}\right)
\eqno(18)
$$
for some set of eigenvalues $\lambda_1,\lambda_2,~\lambda_3$, and
$$\left(\matrix{A(z)&
                  \cr
                B(z)&
                  \cr
                 C(z)&
                  \cr}\right)~=~T^{-1}\left(\matrix{e^{-\lambda_1}& 0 & 0
                                                \cr
                                             0& e^{-\lambda_2} & 0
                                                 \cr
                                             0& 0& e^{-\lambda_3}
                                                  \cr}\right) T
                     \left(\matrix{\int^z~a(z^\prime)dz^\prime &
                                           \cr
                                   \int^z~b(z^\prime)dz^\prime &
                                           \cr
                                    \int^z~c(z^\prime)dz^\prime &
                                            \cr}\right).
\eqno(19)
$$

Since $Ai(z)={1\over 3}{\sqrt z}\left[I_{-{1\over 3}}
\left({2\over 3}z^{3\over 2}\right)
-I_{1\over 3}\left({2\over 3}z^{3\over 2}\right)\right]$ and
$Ai^\prime(z)=-{1\over 3}z\left[I_{-{2\over 3}}\left({2\over 3}z^{3\over 2}
\right)-I_{2\over 3}\left({2\over 3}z^{3\over 2}\right)\right]$

$$\eqalign{\int~dz&~{{(1+k^\prime z)}\over {(1-kz)^2}}Ai(z)Ai^\prime(z)
\cr
~&=~-{1\over 9}\int~dz~{{(1+c^\prime z)}\over {(1-cz)^2}}z^{3\over 2}
\biggl[I_{-{1\over 3}}\left({2\over 3}z^{3\over 2}\right)I_{-{2\over 3}}
\left({2\over 3}z^{3\over 2}\right)
-I_{1\over 3}\left({2\over 3}z^{3\over 2}\right)
I_{-{2\over 3}}\left({2\over 3}z^{3\over 2}\right)
\cr
&~~~~~~~~~~~~~~~~~~~~~~~~~~~~~~~~~~~~~
-I_{-{1\over 3}}\left({2\over 3}z^{3\over 2}\right)
I_{2\over 3}\left({2\over 3}z^{3\over 2}\right)
+I_{1\over 3}\left({2\over 3}z^{3\over 2}\right)
I_{2\over 3}\left({2\over 3}z^{3\over 2}\right)\biggr]
\cr
\int~{{dz}\over {(1-kz)^2}}&[Ai(z)]^2=
\int~{{dz}\over {(1-kz)^2}}~z\biggl[I_{-{1\over 3}}\left({2\over 3}z^{3\over 2}
\right)I_{-{1\over 3}}\left({2\over 3}z^{3\over 2}\right)-2I_{-{1\over 3}}
\left({2\over 3}z^{3\over 2}\right)I_{1\over 3}\left({2\over 3}z^{3\over 2}
\right)
\cr
&~~~~~~~~~~~~~~~~~~~~~~~~~~~~~~~~~~~~~~~~~~~~~~~~~~~
+I_{1\over 3}\left({2\over 3}z^{3\over 2}\right)
I_{1\over 3}\left({2\over 3}z^{3\over 2}\right)\biggr]
\cr}
\eqno(20)
$$
with $\phi(z)={2\over 3}z^{3\over 2}$.

Substituting this function of $z$ into the matrix $M(z)$ and integrating
gives
$$\int^z~M(z^\prime)dz^\prime=\left(\matrix{
      {3\over 2}(\mu+\nu)ln~z&{2\over 3}z^{3\over 2}&{2\over 3}z^{3\over 2}
                 \cr
   -{2\over 3}z^{3\over 2}&-{3\over 2}(1+2\mu)ln~z+{2\over 9}{1\over 
{\mu-\nu}}z^3
  & {3\over 2}(\mu+\nu)ln~z-{2\over 9}{1\over {\mu-\nu}}z^3
    \cr
   -{2\over 3}z^{3\over 2} & {2\over 9}{1\over {\mu-\nu}}z^3  &
                    {3\over 2}(\mu-\nu-1)ln~z-{2\over 9}{1\over {\mu-\nu}}z^3
       \cr}\right).
\eqno(21)
$$
The eigenvalues are roots of the cubic equation
$$\eqalign{\lambda^3~&+~3~ln~z~\lambda^2~+~\left[{2\over 3}z^3ln~z-
{9\over 4}(3\mu^2+\nu^2+\mu+\nu-1)ln^2z\right]\lambda~+~
{4\over 3}(\mu+\nu+1)z^3ln~z
\cr
~&-~(\mu+\nu)z^3ln^2z~+~{{27}\over 8}(\mu+\nu)(1+2\mu)(\mu-\nu-1)ln^3z~=~0.
\cr}
\eqno(22)
$$ 
Defining the coefficients $\alpha,~\beta,~\gamma$ by using the standard
form of the cubic equation $\lambda^3+\alpha\lambda^2+\beta\lambda+\gamma=0$,
it follows that
$$\eqalign{p~&=~-{9\over 4}\left(3\mu^2+\nu^2+\mu+\nu+{1\over 3}\right)ln^2z
~+~{2\over 3}z^3ln~z
\cr
q~&=~{9\over 8}\left(6\mu^3-6\mu\nu^2+3\mu^2-\nu^2-6\mu\nu-\mu-\nu-{2\over 9}
\right)~ln^3z~+~{1\over 3}(\mu+\nu+2)z^3ln^2z.
\cr}
\eqno(23)
$$

When $\mu=-{1\over 3}$ and $\nu=-{2\over 3}$,
$$Q~=~\left({p\over 3}\right)^3+\left({q\over 2}\right)^2
~=~{8\over {729}}z^9ln^3z~+~{2\over {81}}z^6ln^4z~+~
{1\over {216}}z^3ln^5z~-~{1\over {1728}}ln^6z
\eqno(23)
$$
and $Q=0$ when $z=1$ or if ${8\over {81}}w^3+{2\over 9}w^2+{1\over {24}}w
-{1\over {192}}=0$ with $w={{z^3}\over {ln~z}}$.  Since this cubic
equation has a single positive root at $w\doteq 0.08516$, $Q>0$ for all
$z>1$.  If $w< -2.12470$, then $Q<0$; if $-2.12469< w < -0.29559$, then $Q>0$; 
and if $-0.29558< w < 0$, then $Q<0$.   The eigenvalues are
$$\eqalign{\lambda_1~&=~\biggl[-{1\over 6}z^3ln^2z+\left[{8\over {729}}
z^9ln^3z
+{2\over {81}}z^6ln^4z+{1\over {216}}z^3ln^5z
-{1\over {1728}}ln^6\right]^{1\over 2}\biggr]^{1\over 3}
\cr
&~~~~+~\biggl[-{1\over 6}z^3ln^2z-\left[{8\over {729}}z^9ln^3z+{2\over {81}}
z^6ln^4z+{1\over {216}}z^3ln^5z-{1\over {1728}}ln^6z\right]^{1\over 2}
\biggr]^{1\over 3}
\cr
\lambda_2~&=~-{1\over 2}\bigg\{\biggl[-{1\over 6}z^3ln^2z+
\left[{8\over {729}}z^9ln^3z+{2\over {81}}z^6ln^4z+{1\over {216}}z^3ln^5z
-{1\over {1728}}\right]^{1\over 2}\biggr]^{1\over 3}
\cr
&~~~~+~\biggl[-{1\over 6}z^3ln^2z
-\left[{8\over {729}}z^9ln^3z+{2\over {81}}z^6ln^4z
+{1\over {216}}z^3ln^5z-{1\over {1728}}ln^6z\right]^{1\over 2}
\biggr]^{1\over 3}\bigg\}
\cr
&~+~{{i{\sqrt 3}}\over 2}\biggl\{\biggl[-{1\over 6}z^3ln^2z+
\left[{8\over {729}}z^9ln^3z+{2\over {81}}z^6ln^4z+{1\over {216}}z^3ln^5z
-{1\over {1728}}ln^6\right]^{1\over 2}\biggr]^{1\over 3}
\cr
&~~~~-~\biggl[-{1\over 6}z^3ln^2z-\left[{8\over {729}}z^9ln^3z
+{2\over {81}}z^6ln^4z+{1\over {216}}z^3ln^5z
-{1\over {1728}}ln^6z\right]^{1\over 2}\biggr]^{1\over 3}\biggr\}
\cr
\lambda_3~&=~-{1\over 2}\biggl\{\biggl[-{1\over 6}z^3ln^2z+
\left[{8\over {729}}z^9ln^3z+{2\over {81}}z^6ln^4z+{1\over {216}}z^3ln^5z
-{1\over {1728}}\right]^{1\over 2}\biggr]^{1\over 3}
\cr
&~~~~+~\biggl[-{1\over 6}z^3ln^2z
-\left[{8\over {729}}z^9ln^3z+{2\over {81}}z^6ln^4z
+{1\over {216}}z^3ln^5z-{1\over {1728}}ln^6z\right]\biggr]^{1\over 3}
\biggr\}
\cr
&~-~{{i{\sqrt 3}}\over 2}\biggl\{\biggl[-{1\over 6}z^3ln^2z+
\left[{8\over {729}}z^9ln^3z+{2\over {81}}z^6ln^4z+{1\over {216}}z^3ln^5z
-{1\over {1728}}ln^6\right]^{1\over 2}\biggr]^{1\over 3}
\cr
&~~~~-~\biggl[-{1\over 6}z^3ln^2z
-\left[{8\over {729}}z^9ln^3z+{2\over {81}}z^6ln^4z
+{1\over {216}}z^3ln^5z-{1\over {1728}}ln^6z\right]\biggr]^{1\over 3}
\biggr\}.
\cr}
\eqno(25)
$$

Similarly, the following expressions are obtained for $p$, $q$ and $Q$ for 
the other values of $\mu$ and $\nu$
$$\eqalign{p~&=~{2\over 3}z^3ln~z~-~{7\over 4}ln^2z~~~~~~~~~~~~~~~~~~~~~~~~
~~~~~~~~~~~~~~~~~~~~~~~~\mu={1\over 3},~\nu=-{2\over 3}
\cr
q~&=~{5\over 9}z^3ln^2z~+~{3\over 2}ln^3z
\cr
Q~&=~{8\over {729}}z^3ln^3z~-~{1\over {108}}z^6ln^4z~+~{{139}\over {216}}
z^3ln^5z~+~{{629}\over {1728}}ln^6z
\cr
p~&=~{2\over 3}z^3ln~z~-~{{13}\over 4}ln^2z~~~~~~~~~~~~~~~~~~~~~~~~~~~~~~~~~
~~~~~~~~~~~~~~~\mu=-{1\over 3},~\nu={2\over 3}
\cr
q~&=~{3\over 2}ln^3z~+~{7\over 9}z^3ln^2z
\cr
Q~&=~{8\over {729}}z^9ln^3z~-~{1\over {108}}z^6ln^4z~+~{{295}\over
{216}}z^3ln^5z~-~{{1225}\over {1728}}ln^6z
\cr}
$$
$$\eqalign{
p~&=~{2\over 3}z^3ln~z~-~{{19}\over 4}ln^2z~~~~~~~~~~~~~~~~~~~~~~~~~~~~~~~~
~~~~~~~~~~~~~~~~\mu={1\over 3},~\nu={2\over 3}
\cr
q~&=~z^3ln^2z~-~{{15}\over 4}ln^3z
\cr
Q~&=~{8\over {729}}z^9ln^3z~-~{5\over {324}}z^6ln^4z~-~{{11}\over {54}}
z^3ln^5z~-~{{49}\over {108}}ln^6z.
\cr}
\eqno(26)
$$\
The transformation matrix $T$ is equal to
$$\eqalign{T~&=~\left(\matrix{v_{11}& v_{21} & v_{31}
                         \cr
                    v_{12} & v_{22} & v_{32}
                          \cr
                    v_{13} & v_{23} & v_{33}
                          \cr}
                              \right)
\cr
&~~~\left(\matrix{v_{i1}&
                     \cr
                   v_{i2}&
                     \cr
                   v_{i3}&
                      \cr}\right)={1\over {{\cal N}_i}}
\left(\matrix{
-{2\over 3}z^{3\over 2}
{{\left[\lambda_i+{3\over 2}(1+3\mu+\nu)ln~z-{4\over 9}{1\over
{\mu-\nu}}z^3\right]}\over
{\left[{4\over 9}z^3+\left[{2\over 9}{1\over {\mu-\nu}}z^3-{3\over 2}
(\mu+\nu)
ln~z\right]\left[\lambda_i-{3\over 2}(\mu+\nu)ln~z\right]\right]}}&
\cr
1&
\cr
-{{{4\over 9}z^3+\left[\lambda_i+{3\over 2}(1+2\mu)ln~z-{2\over 9}{1\over
{\mu-\nu}}z^3\right]\left[\lambda_i-{3\over 2}(\mu+\nu)ln~z\right]}
\over
{\left[{4\over 9}z^3+\left[{2\over 9}{1\over {\mu-\nu}}z^3-{3\over 2}
(\mu+\nu)
ln~z\right]\left[\lambda_i-{3\over 2}(\mu+\nu)ln~z\right]\right]}}&
\cr}
\right)
\cr}
\eqno(27)
$$
where ${\cal N}_i$ is a normalization factor, so that
$$\eqalign{A(z)~&=~(e^{-\lambda_1}v_{11}^\ast v_{11}+e^{-\lambda_2}
v_{12}^\ast v_{12}+e^{-\lambda_3}v_{13}^\ast v_{13})
\int^z~a(z^\prime)dz^\prime
\cr
&~~~~+(e^{-\lambda_1}v_{11}^\ast v_{21}+e^{-\lambda_2}v_{12}^\ast v_{22}
+e^{-\lambda_3}v_{13}^\ast v_{23})\int^z~b(z^\prime)dz^\prime
\cr
&~~~~+(e^{-\lambda_1}v_{11}^\ast v_{31}+e^{-\lambda_2}v_{12}^\ast v_{32} 
+e^{-\lambda_3}v_{13}^\ast v_{31})\int^z~c(z^\prime)dz^\prime
\cr}
$$
$$\eqalign{
B(z)~&=~(e^{-\lambda_1}v_{21}^\ast v_{11}+e^{-\lambda_2}
v_{22}^\ast v_{12}+e^{-\lambda_3}v_{23}^\ast v_{13})\int^z~a(z^\prime)
dz^\prime
\cr
&~~~~+(e^{-\lambda_1}v_{21}^\ast v_{21}+e^{-\lambda_2}v_{22}^\ast v_{22}
+e^{-\lambda_3}v_{23}^\ast v_{23})\int^z~b(z^\prime)dz^\prime
\cr
&~~~~+(e^{-\lambda_1}v_{21}^\ast v_{31}+e^{-\lambda_2}v_{22}^\ast v_{32} 
+e^{-\lambda_3}v_{23}^\ast v_{31})\int^z~c(z^\prime)dz^\prime
\cr
C(z)~&=~(e^{-\lambda_1}v_{31}^\ast v_{11}+e^{-\lambda_2}
v_{32}^\ast v_{12}+e^{-\lambda_3}v_{33}^\ast v_{13})\int^z~a(z^\prime)
dz^\prime
\cr
&~~~~+(e^{-\lambda_1}v_{31}^\ast v_{21}+e^{-\lambda_2}v_{32}^\ast v_{22}
+e^{-\lambda_3}v_{33}^\ast v_{23})\int^z~b(z^\prime)dz^\prime
\cr
&~~~~+(e^{-\lambda_1}v_{31}^\ast v_{31}+e^{-\lambda_2}v_{32}^\ast v_{32} 
+e^{-\lambda_3}v_{33}^\ast v_{31})\int^z~c(z^\prime)dz^\prime.
\cr}
\eqno(28)
$$
For the first integral in equation (20)
$$\eqalign{\int^z~a(z^\prime)dz^\prime~&=~{1\over {k^{5\over 2}}}
\biggl[{\sqrt {kz}}{{3-2kz}\over {1-kz}}-sinh^{-1}\left[\left({{kz}\over
{1-kz}}\right)^{1\over 2}\right]\biggr]+{2\over {k^{5\over 2}}}
ln~\left({{(1-kz)^{1\over 2}}\over {1+{\sqrt {kz}}}}\right)
+{2\over 3}{{k^\prime}\over {k^2}}z^{3\over 2}
\cr
&~~~+~{{k^\prime}\over {k^{7\over 2}}}\biggl({\sqrt {kz}}
{{5-4kz}\over {1-kz}}-sinh^{-1}
\left[\left({{kz}\over {1-kz}}\right)^{1\over 2}
\right]\biggr)+4{{k^\prime}\over {k^{7\over 2}}}
ln~\left({{(1-kz)^{1\over 2}}\over {1+{\sqrt {kz}}}}\right)
\cr
\int^z~b(z^\prime)dz^\prime~&=~{2\over 3}(\mu+\nu)z^{3\over 2}-{8\over {81}}
{1\over {\mu-\nu}}z^{9\over 2}
\cr
\int^z~c(z^\prime)dz^\prime~&=~-{8\over {81}}{1\over {\mu-\nu}}z^{9\over 2}.
\cr}
\eqno(29)
$$
The equality of the indices $\mu,~\nu$ in the second integral gives rise to
a singularity in the matrix $\int^z~M(z^\prime)dz^\prime$, which can
avoided by using the identity $I_{\nu-1}(z)-I_{\nu+1}(z)={{2\nu}\over z}
I_\nu(z)$, to obtain an integral containing the products of modified Bessel 
functions different indices.  The correction to the standard wave function
can be obtained through the substitution $z\to -z$ in the integrals (20).

The contribution of the additional integral in equation (12) to the wave 
function includes $Ai(-z)$ so that the functional dependence of the 
normalization factor $Ai(-z_0)+C_1{{e^{-\Phi}}\over {g_4^2}}Ai(-z_0)
+{{e^{-\Phi}}\over {g_4^2}}f(z_0)$ is divisible by $Ai(-z_0)$.  Given
that the normalization factor contains
$Ai\left(K\left({{36}\over V}\right)^{2\over 3}\right)$, the conclusions
concerning the feasibility of predicting an inflationary universe through
the quantum cosmological wave function are unaltered by the addition of the
higher-order curvature terms.

In the Planck era, the higher-order curvature terms in the perturbative
expansion of the string effective action have approximately the same
magnitude as the Ricci scalar, and similarly it is inappropriate to use a
truncated form of a series expansion in ${{e^{-\Phi}}\over {g_4^2}}$
of the Wheeler-De Witt equation.  Instead, a closed-form sixth order
differential equation, which can be obtained by including the conjugate
momentum to derivative of the scale factor $P_{\dot a}$, and then
using the Ostrogradski method to define the Hamiltonian for the 
higher-derivative theory, can be used to define the quantum cosmological
wave function in the initial era.  
Given the one-dimensional action
$$I~=~\int~dt~\left[(6a^2{\ddot a}+6a{\dot a}^2+6aK)+{1\over 2}a^3{\dot \Phi}^2
~+~6{{e^{-\Phi}}\over {g_4^2}}{\ddot a}({\dot a}^2+K)\right]
\eqno(30)
$$
and the conjugate momenta
$$\eqalign{P_a~&=~{{\partial L}\over {\partial {\dot a}} }-{d\over {dt}}
\left({{\partial L}\over {\partial {\ddot a}}}\right)~=~6{{e^{-\Phi}}\over
{g_4^2}}{\dot \Phi}({\dot a}^2+K)
\cr
P_{\dot a}~&=~{{\partial L}\over {\partial {\ddot a}}}
~=~6\left(a^2+{{e^{-\Phi}}\over {g_4^2}}({\dot a}^2+K)\right)
\cr
P_\Phi~&=~{{\partial L}\over {\partial {\dot \Phi}}}~=~a^3 {\dot \Phi}
\cr}
\eqno(31)
$$
the Hamiltonian is
$$\eqalign{
H~&=~P_a{\dot a}~+~P_{\dot a} {\ddot a}~+~P_\Phi{\dot \Phi}~-~L
\cr
&=~-6a({\dot a}^2+K)~+~6{{e^{-\Phi}}\over {g_4^2}}{\dot \Phi}
({\dot a}^2+K){\dot a}~+~{1\over 2}a^3{\dot \Phi}^2
\cr
&=~-g_4^2P_\Phi^{-1}e^\Phi a^4P_a~+~{1\over {2a^3}}P_\Phi^2
~+~\left[{{g_4^2}\over 6}P_\Phi^{-1}e^\Phi P_a^2 
a^3P_a~-~KP_a^2\right]^{1\over 2}
\cr}
\eqno(32)
$$
and the pseudo-differential equation $H\Psi=0$ can be transformed into
the partial differential equation \footnote{*}{This equation differs from
the equation (11) in reference [14] by a derivative with respect to $\Phi$.  
Imposing an additional constraint on the wave function, the derivative term 
can be eliminated through the addition of an extra term in the Lagrangian.}
$$\eqalign{-{{g_4^2}\over 6}&e^{-\Phi}\left(a^3{{\partial^4\Psi}\over
{\partial a^3 \partial\Phi}}+6a^2{{\partial^3\Psi}\over {\partial a^2
\partial \Phi}}\right)~+~Ke^{-2\Phi}\left({{\partial^4\Psi}\over
{\partial a^2 \partial\Phi^2}}-{{\partial^3\Psi}\over {\partial a^2
\partial\Phi}}\right)
\cr
&=~a^4g_4^4\left[4a^3{{\partial \Psi}\over {\partial a}}+a^4
{{\partial^2\Psi}\over {\partial a^2}}\right]~+~ag_4^2e^{-\Phi}
\left({{\partial^4\Psi}\over {\partial a^2 \partial \Phi^2}}~+~
{{\partial^3\Psi}\over {\partial a \partial\Phi^2}}\right)
\cr
&~+{3\over 2}g_4^2e^{-\Phi}\left(a{{\partial^2\Psi}\over
{\partial a \partial \Phi}}~-~{{\partial^3\Psi}\over {\partial \Phi^3}}\right)
~+~{1\over {4a^6}}e^{-\Phi}{\partial\over {\partial \Phi}}
\left(e^{-\Phi}{{\partial^5\Psi}\over {\partial \Phi^5}}\right).
\cr}
\eqno(33)
$$

By including a potential term in the Lagrangian and discarding terms 
containing derivatives of $\Psi$ and $V(\Phi)$ with respect to $\Phi$, the 
following equation is obtained
$${{a^2g_4^2}\over 6}{{d^3\Psi}\over {da^3}}+ag_4^2(1+a^6e^\Phi)
{{d^2\Psi}\over {da^2}}+[g_4^2(1+a^6e^\Phi)-2g_4^2a^6V(\Phi)]
{{d\Psi}\over {da}}-3g_4^2a^5V(\Phi)\Psi=0
\eqno(34)
$$

The corrected wave function in the inflationary epoch can be matched with the 
solution to a third-order partial differential equation in the initial era 
along a boundary which also must be determined by setting the derivatives
up to second order in $a$ to be equal.

While the wave function $\Psi(a,\Phi)$ has been used to establish 
whether the curvature-dependence defined by the potential favours inflation, 
to determine the most probable path in minisuperspace $\{a(t),\Phi(t)\}$, 
it is preferable to consider the partition function
$$Z~=~\int~e^{-I[a(t),\Phi(t)]}~d[a(t)]d[\Phi(t)]
\eqno(35)
$$
which is extremized at the classical solutions
$${{\delta Z}\over {\delta a}}\bigg\vert_{a_{cl.}}~=~
{{\delta Z}\over {\delta \Phi}}\bigg\vert_{\Phi_{cl.}}~=~0.
\eqno(36)
$$
The expectation values $\langle a(t)\rangle$ and $\langle \Phi(t)\rangle$
based on the action $I$ are

$$\eqalign{\langle a(t)\rangle~&
=~{{\int~a(t)e^{-I[a(t),\Phi(t)]}d[a(t)]d[\Phi(t)]}
\over {\int~e^{-I[a(t),\Phi(t)]}~d[a(t)]d[\Phi(t)]}}
\cr
&\simeq~{{\int~a(t)e^{-\left[{{\delta^2 I}\over 
{\delta a(t)^2}}(\delta a(t))^2~+~2{{\delta^2 I}\over {\delta a(t)
\delta \Phi(t)}}\delta a(t)\delta \Phi(t)~+~{{\delta^2 I}\over
{(\delta \Phi(t))^2}}(\delta \Phi(t))^2\right]}~d[a(t)]d[\Phi(t)]}\over
{\int~e^{-\left[{{\delta^2 I}\over 
{\delta a(t)^2}}(\delta a(t))^2~+~2{{\delta^2 I}\over {\partial a(t)
\partial \Phi(t)}}\delta a(t)\delta \Phi(t)~+~{{\delta^2 I}\over
{(\delta \Phi(t))^2}}(\delta \Phi(t))^2\right]}~d[a(t)]d[\Phi(t)]}}
\cr
\langle \Phi(t)\rangle~&=~{{\int~\Phi(t)~e^{-I[a(t),\Phi(t)]}d[a(t)]
                                                  d[\Phi(t)]}
                           \over {\int~e^{-I[a(t),\Phi(t)]}~d[a(t)]d[\Phi(t)]}}
\cr
&\simeq~{{\int~\Phi(t)~e^{-\left[{{\delta^2 I}\over 
{\delta a(t)^2}}(\delta a(t))^2~+~2{{\delta^2 I}\over {\delta a(t)
\delta \Phi(t)}}\delta a(t) \delta \Phi(t)~+~{{\delta^2 I}\over
{(\delta \Phi(t))^2}}(\delta \Phi(t))^2\right]}~d[a(t)]d[\Phi(t)]}\over
{\int~e^{-\left[{{\delta^2 I}\over {\delta a(t)^2}}(\delta a(t))^2~
+~2{{\delta^2 I}\over {\delta a(t)\delta \Phi(t)}}\delta a(t)\delta 
\Phi(t)~+~{{\delta^2 I}\over {(\delta \Phi(t))^2}}
(\delta \Phi(t))^2\right]}~d[a(t)]d[\Phi(t)]}}
\cr}.
\eqno(37)
$$  

Since
$$\eqalign{ {{\delta^2 I}\over {\delta a^2}}~&=~
  \int~dt~\left[{{\partial^2L}\over {\partial a^2}}~-{3\over 2}~{d\over {dt}}
\left({{\partial^2L}\over {\partial a \partial {\dot a}}}\right)~+~{1\over 2}
{{d^2}\over {dt^2}}\left({{\partial^2L}\over {\partial {\dot a}^2}}\right)
\right]
\cr
{{\delta^2 I}\over {\delta \Phi^2}}~&=~
\int~dt~\left[{{\partial^2L}\over {\partial \Phi^2}}~-{3\over 2}~{d\over {dt}}
\left({{\partial^2L}\over {\partial \Phi \partial {\dot \Phi}}}\right)~+~
{1\over 2}{{d^2}\over {dt^2}}\left({{\partial^2L}\over 
{\partial {\dot \Phi}^2}}\right)\right]
\cr}
\eqno(38)
$$
and the equation of motion for $a(t)$ implies that
${{\partial L}\over {\partial a}}-{d\over {dt}}\left({{\partial L}\over
{\partial {\dot a}}}\right)=\sum_{n\ge 1}~f_n(a,{\dot a},\Phi,{\dot \Phi})
(a(t)-a_{cl}(t))^n$, 
$$\eqalign{
{\partial\over {\partial\Phi}}\left({{\partial L}\over {\partial a}}~-~
{d\over {dt}}\left({{\partial L}\over {\partial {\dot a}}}\right)\right)
&~=~\sum_{n\ge 1}~{{\partial f_n(a,{\dot a},\Phi,{\dot \Phi})}
\over {\partial \Phi}}(a(t)-a_{cl}(t))^n
\cr
{\partial\over {\partial {\dot\Phi}}}\left({{\partial L}\over {\partial a}}~-~
{d\over {dt}}\left({{\partial L}\over {\partial {\dot a}}}\right)\right)
&~=~\sum_{n\ge 1}~{{\partial f_n(a,{\dot a},\Phi,{\dot \Phi})}
\over {\partial {\dot \Phi}}}(a(t)-a_{cl}(t))^n
\cr}
\eqno(39)
$$
both vanish when $a(t)=a_{cl}(t)$.  Similarly,
${\partial\over {\partial a}}\left({{\partial L}\over {\partial\Phi}}
~-~{d\over {dt}}\left({{\partial L}\over {\partial {\dot \Phi}}}\right)\right)
~=~{\partial\over {\partial {\dot a}}}\left({{\partial L}\over {\partial\Phi}}
~-~{d\over {dt}}\left({{\partial L}\over {\partial {\dot \Phi}}}\right)
\right)~=~0$ 
when $\Phi(t)=\Phi_{cl}(t)$.  From the equation (4), 
$$\eqalign{ {{\partial^2L}\over {\partial a^2}}~-~{3\over 2}{d\over {dt}}
\left({{\partial^2L}\over {\partial a \partial{\dot a}}}\right)~+~
{1\over 2}{{d^2}\over {dt^2}}
\left({{\partial^2L}\over {\partial {\dot a}}}\right)^2
~&=~3a{\dot \Phi}^2-6aV(\Phi)~+~12 {\ddot a}~+~
{6\over {g_4^2}}\biggl(-3e^{-\Phi}{\dot \Phi}{\ddot \Phi}{\dot a}
\cr
&~~~-~2e^{-\Phi}{\dot \Phi}^2{\ddot a}+~2e^{-\Phi}{\ddot \Phi}{\ddot a}
~+~e^{-\Phi}{{d^3\Phi}\over {dt^3}}{\dot a}~+~
e^{-\Phi}{\dot \Phi}{{d^3a}\over {dt^3}}
\cr
&~~~~+~e^{-\Phi}{\dot \Phi}^3{\dot a}
\biggr)
\cr
{{\partial^2L}\over {\partial \Phi^2}}~-~{3\over 2}{d\over {dt}}
\left({{\partial^2L}\over {\partial \Phi \partial {\dot \Phi}}}\right)
~+~{1\over 2}{{d^2}\over {dt^2}}\left({{\partial^2L}\over {\partial
{\dot\Phi}^2}}\right)~&=~-a^3V^{\prime\prime}(\Phi)~+~2 {{e^{-\Phi}}
\over {g_4^2}}{\dot \Phi}{\dot a}({\dot a}^2+3K)
\cr
&~~~~+~3{{e^{-\Phi}}\over {g_4^2}}{\dot \Phi}{\dot a}({\dot a}^2+3K)
-9{{e^{-\Phi}}\over {g_4^2}}{\ddot a}({\dot a}^2+K)
\cr
&~~~~+~{9\over 4}(2a {\dot a}^2+3a^2{\ddot a}).
\cr}
\eqno(40)
$$
and substituting the approximate solutions to the equations of motion for
$a(t)$ and $\Phi(t)$ [10], based on the heterotic string potential,
\footnote{*}{The opposite sign to the usual convention must be chosen to 
obtain a positive value for 
\hfil\break
\phantom{....}$V(\Phi=0)$.}

$$\eqalign{a(t)~&=~a_0e^{\lambda t}
\cr
\Phi(t)~&\sim~ln\left\vert~{{27}\over {4C}}{{h^2e^{-3\sigma_0}}\over
{b_0^3 g_4^4}}~cosh({\sqrt C}(t-t_0))\right\vert
\cr
18\lambda^2~&-~{3\over 2}C~+~{3\over {16}}g_4^2e^{-3\sigma_0}k^2e^D~=~0  
\cr
c+h~&=~ke^{-{{\sqrt C}\over 2}t}
\cr}
\eqno(41)
$$
gives 
$$\eqalign{{{\partial^2L}\over {\partial a^2}}-{3\over 2}{d\over {dt}}
\left({{\partial^2L}\over {\partial a\partial{\dot a}}}\right)+
{1\over 2}{{d^2}\over {dt^2}}
\left({{\partial^2L}\over {\partial {\dot a}^2}}\right)
&{\longrightarrow\atop {t\to \infty}} 
\left[3a_0(1+4\lambda^2)+
{{g_4^2e^{-3\sigma_0}}\over {16}}k^2\right]e^{\lambda t}+
{\cal O}(e^{(\lambda-{\sqrt C})t})>0 
\cr
{{\partial^2L}\over {\partial \Phi^2}}-{3\over 2}{d\over {dt}}
\left({{\partial^2L}\over {\partial \Phi \partial {\dot \Phi}}}\right)
+{1\over 2}{{d^2}\over {dt^2}}&\left({{\partial^2L}\over {\partial
{\dot\Phi}^2}}\right){\longrightarrow\atop {t\to \infty}}
{9\over 4}a_0^3\lambda^2(2+3\lambda^2)e^{3\lambda t}+
{\cal O}\left(e^{(3\lambda-{\sqrt C})t}\right)>0.
\cr}
\eqno(42)
$$
To second order, the probability distributions about the
classical paths in minisuperspace would be Gaussian, and average
values $\langle a(t)\rangle,~\langle \Phi(t)\rangle$ equal $a_{cl}(t),~
\Phi_{cl}(t)$.  A shift in the expectation values $\langle a(t)\rangle,~
\langle \Phi(t) \rangle$ only arises at third order in the expansion
of the functional integrals in equation (36).

\vskip 10pt

\centerline{\bf Acknowledgements}

\noindent
This work has been supported by an ARC Small Grant.  Much of the discussion 
of the renormalizability properties of the action, the exponential prefactor 
in the no-boundary wave function, the differential equation for the wave 
function in the Planck era and the average paths in the minisuperspace, is 
based on research completed at the Institut f{\"u}r Mathematik.

\vskip 10pt

\centerline{\bf References}
\item{[1]} E. Elizalde, A. G. Jacksenaev, S. D. Odintsov and I. L. Shapiro,
Phys. Lett. ${\underline{328B}}$ (1994) 297-306 
\item{[2]} A. O. Barvinsky, Nucl. Phys. ${\underline{B561}}$ (1999) 159-187
\item{[3]} P. W. Higgs, Nuovo Cimento ${\underline{11}}$ (1959) 816-820
\item{[4]} G. V. Bicknell, J. Phys. A ${\underline{7}}$ (1974) 341-345
\item{[5]} B. Whitt, Phys. Lett. ${\underline{B145}}$ (1984) 176-178
\item{[6]} G. Magnano, M. Ferraris and M. Francaviglia, Gen. Rel. Grav. 
${\underline{19}}$ (1987) 465-479 
\item{[7]} A. Jakubiec and J. Kijowski, Phys. Rev. ${\underline{D37}}$
(1988) 1406-1409
\item{[8]} M. Ferraris, M. Francaviglia and G. Magnano, Class. Quantum
Grav. ${\underline{5}}$ (1988) L95
\item{[9]} K. Stelle, Phys. Rev. ${\underline{D16}}$ (1977) 953-969
\item{[10]} S. Davis and H. C. Luckock, Phys. Lett. ${\underline{485B}}$ 
(2000) 408-421
\item{[11]} R. Graham and H. Luckock, Phys. Rev. ${\underline{D49}}$(10)
(1994) 4981-4984
\item{[12]} N. Kontoleon and D. W. Wiltshire, Phys. Rev. ${\underline{D59}}$
(1999) 063513:1-8
\item{[13]} D. Wiltshire, Gen. Rel. Grav. ${\underline{32}}$ (2000) 515-528
\item{[14]} ${\underline{Handbook~of~Mathematical~Functions}}$, eds.
M. Abramowitz and I. A. Stegun (New York: Dover Publications, Inc., 1970)
\item{[15]} N. Sonine, Math. Ann. ${\underline{XVI}}$ (1880) 1-80 
\item{[16]} S. Davis, Gen. Rel. Grav. ${\underline{32}}$(3) (2000) 541-551

\end